\documentclass[aps,twocolumn,
showkeys,showpacs,nofootinbib,byrevtex]{revtex4-1}
\usepackage{amsmath,amsfonts,amssymb,amscd,amsxtra,amsthm}
\usepackage{graphicx}
\usepackage{bm}
\usepackage{epsfig}

\begin{document}

\title{Features of $\omega$ photoproduction off nucleon target
at forward angles : Dominance of $\pi$ exchange with Regge cuts
and scaling of differential cross sections}

\author{Byung-Geel Yu}%
\email{bgyu@kau.ac.kr}%
\affiliation{ Research Institute of Basic Science, Korea Aerospace
University, Goyang, 10540, Korea}

\author{Tae Keun Choi}%
\email{tkchoi@yonsei.ac.kr}%
\affiliation{ Dept. of Physics, Yonsei University, Wonju, 220-710,
Korea}

\author{Kook-Jin Kong}%
\email{kong@kau.ac.kr}%
\affiliation{Research Institute of Basic Science, Korea Aerospace
University, Goyang, 10540, Korea}


\begin{abstract}
We investigate photoproduction of $\omega$ off a nucleon target
$\gamma N\to\omega N$ by using a Reggeized model where
$\pi(135)+\sigma(500)+f_1(1285)+f_2(1270)$+Pomeron exchanges are
included in the $t$-channel for the reaction at forward angles.
The reaction mechanism at low energy is featured by the dominance
of the $\pi$ exchange with the absorptive cuts introduced to
modulate the pion contribution to both $\gamma p\to \omega p$ and
$\gamma n\to \omega n$ reactions. Necessity of $\sigma$ exchange
is illustrated in the analysis of the cross section from natural
parity exchanges. Cross sections for differential, total, and spin
density matrix with parity and beam polarization asymmetries
are reproduced and compared
with existing data on $\gamma p\to \omega p$. Scaled differential
cross sections of Jefferson Lab data on $\gamma p\to \omega p$ are
investigated at the production angle $\theta=90^\circ$ with the
nonlinear trajectories for a saturation. Differential and total
cross sections for the $\gamma n\to\omega n$ reaction are analyzed
to compare with recent experimental data at the CBELSA/TAPS
Collaboration. An application to the $\gamma p\to\omega\Delta^+$
reaction is presented to demonstrate the significance of the $\pi$
exchange in the total and differential cross sections.
\end{abstract}

\maketitle

\section{introduction}

Photoproductions of lighter vector mesons, $\rho^0$, $\omega$, and
$\phi$ are important to study hadron reactions because they
provide information to understand the diffractive scattering that
leads to a discovery of the Pomeron \cite{laget}.
Among these reactions photoproduction of $\omega$, in particular,
shows a feature quite distinctive from others in the low energy
region by the special role of the $\pi$ exchange. Given the decay
width $\omega\to\pi^0\gamma$ especially large, the role of $\pi$
exchange appears to be prominent more  than the cases in the
$\rho^0$ and $\phi$ processes near threshold \cite{friman}.
Furthermore, as an application to nuclear reactions to observe
vector meson properties in nuclear medium \cite{chudakov}, the
recent 12 GeV upgrade of the Jefferson Lab facilities with the
GlueX detector draws attention because the knowledge about
$\omega$ photoproduction becomes even more important to understand
QCD dynamics through the color transparency in the nuclear
photoproduction of $\omega$ \cite{chudakov,sibirtsev-ct}.
Therefore, all these topics need for a comprehensive study of
$\omega$ photoproduction on the nucleon target through the
systematic approach to the analysis of the empirical data
\cite{sibirtsev}.

In this work we elaborate to construct a model for $\omega$
photoproduction at forward angles and attempt to describe the
production mechanism throughout the reaction energy from threshold
to hundreds of GeV region where only the Pomeron exchange is
prevailing. For this purpose, we utilize the Regge model of Ref.
\cite{kong2} where the $t$-channel meson exchanges
$\pi+\sigma+f_2$ are included without either fit parameters for
coupling strength or form factors to cutoff divergences. Within
the Reggeized framework the roles of lighter mesons $\pi$ and
$\sigma$ are worth investigating, because the former exchange with
the large couplings of $\gamma\pi\omega$ and $\pi NN$ gives the
contribution too excessive to agree with the observed cross
section and the latter exchange plays the role more important than
the $\pi$ exchange in the peaking of $\rho^0$ \cite{cano,friman}
and $\phi$ cross sections near threshold \cite{kong2}.

To clarify the roles of these nonresonant meson exchanges the
measurement of the $\gamma p\to\omega p$ process at the SLAC/LBL
Collaborations in Ref. \cite{ballam} is useful because the
reaction cross section is provided in a separate manner into the
natural and the unnatural parity exchanges respectively at photon
energies 2.8, 4.7, and 9.3 GeV. This enables us to investigate
meson exchanges of different parities independently, because they
do not interfere with each other.
More recently, experiments on $\omega$ photoproduction are
extended to include the reaction at the deuteron target and the
data on the reaction off a neutron target from the CBELSA/TAPS and
GRAAL Collaborations \cite{dietz,wilson}  are available for photon
energies up to 2 GeV.
From the isospin symmetry, therefore, photoproduction of $\omega$
off the neutron target could provide further constraint on the
different role of the $\pi$ exchange in the isovector channel.

The scaling of the differential cross section is one of the
interesting phenomena observed in hadron reactions  at wide
angles, or alternatively large $-t$.
According to the dimensional scaling predicted from pQCD
calculation \cite{brodsky}, the reaction around mid angle, i.e.,
the production angle $\theta\approx 90^\circ$, shows a scaling of
cross section as the reaction energy increases
\cite{kong2,zhu,dey}. Thus, quark evidences in photoproduction of
hadrons could be searched for and the measurement of cross section
for $\gamma p\to\omega p$ at the angle around $\theta=90^\circ$ is
also expected to offer an observation of the transition between
hadronic and parton phases in the target nucleon.
In the present framework which is specialized to describe the
reaction at forward angles we make an analysis of the scaled
differential cross sections by virtue of the saturation of the
Regge trajectory valid at large $-t$.

In relation with the above issues it is worth investigating the
reaction $\gamma p\to\omega\Delta^+$ in parallel with the $\omega$
photoproduction on nucleon, because the reaction has not been well
understood yet with few experiments \cite{barber,junkersfeld07}
and theories \cite{clark77}. Together with the $\gamma p\to\omega
p$ reaction the knowledge of the reaction mechanism may help to
analyze nuclear reactions including the study of color
transparency in nuclear photoproduction of $\omega$.

All the topics discussed above are our primary interest in the
present work and this paper is organized as follows. In Sec. II we
construct a model for the meson exchange which is reggeized in the
$t$-channel. Contributions of the natural and unnatural parity
exchanges are investigated, respectively, based on the SLAC/LBL
data as discussed above. The role of $\sigma$ in addition to
$f_2+$Pomeron exchanges is discussed in the natural parity cross
section, while the necessity of the Regge cuts to modulate the
$\pi$ exchange is demonstrated for the $\gamma N\to\omega N$,
which, otherwise, overestimates the unnatural parity cross
section.
Numerical consequence in $\gamma p\to\omega p$ reaction are
presented in Sec. III for differential, total, and scaled
differential cross sections. Spin polarizations including spin
density matrix and beam polarization are analyzed to compare with
data over resonance region. Differential and total cross sections
for  $\gamma n\to\omega n$ process are calculated and given to
compare with recent CBELSA/TAPS data. As an extension of the
current framework, the application to $\gamma p\to\omega \Delta^+$
process is also presented to investigate the relevance of the
$\pi$ exchange with the cut. Section IV contains a summary and
discussion.

\section{ The Regge model}

In this section we construct a photoproduction amplitude for the
reaction
\begin{eqnarray}\label{react}
\gamma(k)+ N(p)\to\omega(q)+ N(p'),
\end{eqnarray}
which is able to describe the Pomeron exchange at high energies,
while in the low and intermediate  energy regions the nonresonant
meson exchanges are to reproduce threshold peaking in the cross
sections measured in experiments. Here $k$, $p$ are the
four-momenta of photon and nucleon in the initial state, and $q$,
$p'$ are the $\omega$ and the final nucleon momenta, respectively.

\subsection{$t$-channel meson Regge poles at forward angles}

Since the $\omega$-meson itself is not allowed to exchange by
charge conjugation, the meson exchanges relevant to the reaction
in Eq. (\ref{react}) are written as
\begin{eqnarray}\label{neutral}
&&{\cal M}={\cal M}_\sigma+{\cal M}_{f_2}+{\cal
M}_\mathbb{P}+{\cal M}_\pi+{\cal M}_{f_1},
\end{eqnarray}
where the scalar meson, tensor meson, and Pomeron exchanges are
\begin{widetext}
\begin{eqnarray}\label{charge0}
&&{\cal M}_\sigma= \frac{g_{\gamma \sigma\omega}}{ m_0}g_{\sigma
NN}(k \cdot q\,\eta^*\cdot\epsilon-\epsilon\cdot q\,\eta^*\cdot
k)\, \bar{u}(p')u(p){\cal R}^{\sigma}(s,t)\,,\\
&&{\cal M}_{f_2}=\Gamma_{\gamma
f_2\omega}^{\beta\rho}(k,q)\Pi_{\beta\rho;\lambda\sigma}(Q)
\bar{u}(p')\Gamma_{f_2NN}^{\lambda\sigma}(p',p)u(p){\cal R}^{f_2}(s,t)\,,\\
&&{\cal M}_\mathbb{P}=12i{e\,\beta_q\beta_{q'}\over f_\omega}
{m_\omega^2\over m_\omega^2-t}\left({2\mu_0^2\over
2\mu_0^2+m_\omega^2-t}\right)e^{-i{\pi\over
2}[\alpha_\mathbb{P}(t)-1]}\left({s\over
4s_0}\right)^{\alpha_\mathbb{P}(t)-1}F_1(t)\bar{u}(p')
(/\kern-6pt{k}\eta^*\cdot\epsilon-\rlap{/}\epsilon\eta^*\cdot
k)u(p)
\end{eqnarray}
for the natural parity ($P=(-1)^J$), and the $\pi$ and $f_1$ axial
vector meson are
\begin{eqnarray} &&{\cal
M}_\pi=i\,\frac{g_{\gamma\pi\omega}}{m_0}g_{\pi NN}
\varepsilon^{\mu\nu\alpha\beta}\epsilon_\mu\eta_\nu^* k_\alpha
q_\beta\,\bar{u}(p')\gamma_5u(p){\cal R}^{\pi}(s,t)\,,\label{pion}\\
&&{\cal M}_{f_1}=i{g_{\gamma f_1\omega}\over m_0^2}g_{f_1
NN}m_\omega^2\epsilon_{\mu\nu\alpha\beta}k^\mu\eta^{\nu*}\epsilon^\alpha
\left(-g^{\beta\lambda}+Q^\beta
Q^\lambda/m^2_{f_1}\right)
\left({1\over1-t/M^2_{A}}\right)^2
\overline{u}(p')\gamma_\lambda\gamma_5 u(p){\cal R}^{f_1}(s,t)\,,
\end{eqnarray}
\end{widetext}
for unnatural parity ($P=-(-1)^J$) exchanges.

The Regge propagator is given by
\begin{eqnarray}
&&{\cal R}^{\varphi}(s,t)={\pi\alpha'_J\times{\rm
phase}\over\Gamma[\alpha_J(t)+1-J]\sin[\pi\alpha_J(t)]}
\left({s\over s_0}\right)^{\alpha_J(t)-J},
\end{eqnarray}
written in the collective form for the $\varphi$ meson of spin-$J$
which stands for all the mesons considered here. $s_0=1$ GeV$^2$.
The phase factor is, in general, taken to be  of the canonical
form, ${1\over 2} [(-1)^J + e^{-i\pi\alpha_J(t)}]$, for each meson
exchange. Photon and vector meson polarizations are denoted by
$\epsilon(k)$ and $\eta^*(q)$ with momenta $k$ and $q$,
respectively. $u(p)$ and $u(p')$ are the spinors for the initial
and final nucleon with the momenta $p$ and $p'$, respectively.
$Q^\mu=(q-k)^\mu$ is the $t$-channel momentum-transfer and the
dimensionful parameter $m_0=1$ GeV.
\\

\subsubsection*{Natural parity exchange}

$\bullet\ $ $\sigma(500)$ $J^{PC}=0^{++}$

Though the isoscalar $\sigma$ meson plays an important role in
reproducing the threshold peak of the $\rho^0$ and $\phi$
reactions \cite{friman,kong2}, it was excluded in previous studies
\cite{titov,ysoh} by the uncertainty in the decay width.
However, the Particle data group (PDG) reports that the partial
widths $\Gamma_{\omega\to\pi^+\pi^-\gamma}\approx 3.6\times
10^{-3}$ and $\Gamma_{\omega\to\pi^0\pi^0\gamma}\approx 6.7\times
10^{-5}$ which are comparable to those of $\rho^0$ and $\phi$  in
magnitude. In practice the two pions can be associated with an
$s$-wave propagation of $\sigma$ as an effective degree of
freedom. Thus, taking $\Gamma_{\omega \to \sigma\gamma}\approx
\Gamma_{\omega \to \pi^+\pi^-\gamma}\approx 30.56$ keV as the
upper limit, we estimate $|g_{\gamma\sigma\omega}|\approx 0.53$,
which is not negligible but rather larger than the radiative
decays of other mesons. Moreover, such an estimate is consistent
with the predictions either
$\Gamma_{\omega\to\sigma\gamma}=16\pm3$, or $33\pm4$ keV from the
chiral effective Lagrangians incorporated with the vector meson
dominance \cite{black}. This yields
$|g_{\gamma\sigma\omega}|\approx 0.22$ or $0.32$, respectively. In
the present calculation, we take the smaller value
$g_{\gamma\sigma\omega}=-0.17$ with the negative sign for the
reason for a better agreement with the natural-parity cross
section as shown in Fig. \ref{fig1} (a). For the $\sigma NN$
coupling constant, we take $g_{\sigma NN}=14.6$ ~\cite{erkol}
close to $\pi NN$ coupling constant to be consistent with the
chiral partner to the $\pi$ in the $\sigma$ model.
\\

$\bullet\ $ $f_2(1270)$ $J^{PC}=2^{++}$

Following the isoscalar $\sigma$ meson, the exchange of spin-2
tensor meson $f_2$ gives the contribution substantial to reproduce
the reaction cross section in intermediate energies up to
$\sqrt{s}\approx 10$ GeV.  The coupling constants of $f_2$
exchange are obtained as $g_{\gamma f_2 \omega}/m_0=0.0376/m_0$
from the partial decay width $\Gamma_{f_2\to\omega\gamma}=27$ keV
by the relativistic quark model prediction \cite{ishida} and
$g^{(1)}_{f_2NN}=6.45$ and $g^{(2)}_{f_2NN}=0$ for the tensor
meson-nucleon couplings from Ref.~\cite{bgyu}.  Details of the
coupling vertices $\Gamma_{\gamma f_2\omega}^{\beta\rho}(k,q)$ and
$\Gamma_{f_2NN}^{\lambda\sigma}(p',p)$ are given in
Ref~\cite{kong2}. The spin-2 projection is given by
\begin{eqnarray}\label{spin2}
\Pi_{(2)}^{\beta\rho;\sigma\lambda}(Q)={1\over
2}(\bar{g}^{\beta\sigma}\bar{g}^{\rho\lambda}
+\bar{g}^{\beta\lambda}\bar{g}^{\rho\sigma})-{1\over
3}\bar{g}^{\beta\rho}\bar{g}^{\sigma\lambda}\,,
\end{eqnarray}
with $\bar{g}^{\mu\nu}(Q)=-g^{\mu\nu}+{Q^\mu Q^\nu}/{m_{a_2}^2}$.
\\

$\bullet\ $ Pomeron  $J^{PC}=1^{-+}$

The concept and practical use of the exchange of vacuum
quantum-numbers, so called the soft Pomeron, is to date
well-established in photoproductions of lighter vector mesons. The
Pomeron exchange of an isoscalar photon-like quantum number
$J^{PC}=1^{-+}$ in the $\omega$ photoproduction plays the role
unique in the slow increase of cross sections up to
$\sqrt{s}\approx 100$ GeV. We use the trajectory
$\alpha_{\mathbb{P}}(t)=0.25\,t+1.08$ and the decay constant
$f_\omega=15.6$ from the SU(3) symmetry,
\begin{eqnarray}
{1\over f_\rho}:{1\over f_\omega}:{1\over f_\phi}=3:1:-\sqrt{2}\,,
\end{eqnarray}
which is consistent with $f_{\phi}=-13.4$ \cite{kong2} and
$f_\rho=5.2=2g_{\rho NN}$ \cite{bgyu-rho}. The parameters for the
quark-Pomeron coupling are given by $\beta_u=\beta_d=2.07$
GeV$^{-1}$, and the cutoff mass $\mu_0^2=1.1$ GeV$^2$. We adopt
the nucleon isoscalar form factor $F_{1}(t)$ as given in Ref.
\cite{kong2}.
\\

\subsubsection*{Unnatural parity exchange}

$\bullet\ $ Pion $J^{PC}=0^{-+}$

To determine the contribution of $\pi$ exchange is important,
because it is crucial to characterize the overall feature of
reaction in the low energy region. The coupling constants $g_{\pi
NN}=\pm13.4$ are used for the $\gamma p$ and $\gamma n$ reactions,
respectively, and $|g_{\gamma\pi\omega}|=0.69$ from the radiative
decay width $\Gamma_{\omega\to\gamma\pi}=0.7$ MeV reported in the
PDG.
However, it is not easy to determine the sign of the radiative
coupling from the unnatural parity cross section, because of the
single dominance of $\pi$ exchange over the contribution of $f_1$
exchange which is negligible. In this work we choose the negative
sign of $g_{\gamma\pi\omega}$ for a fair agreement with
experimental data. In the reggeization, the phase of $\pi$
exchange takes the canonical form, because the exchange-degenerate
partner $b_1$ of negative $C$-parity is absent from the reaction.
Nevertheless, we take the complex phase for the $\gamma p$
reaction, which is consistent with the scaled differential cross
section as will be discussed later. For the $\gamma n$ reaction we
take the constant phase for the $\pi$ Reggeon to agree with
differential cross sections of Ref. \cite{dietz}.
\\

$\bullet\ $ $f_1(1285)$ $J^{PC}=1^{++}$

\begin{table}[t]
\caption{Listed are the physical constants and Regge trajectories
with the corresponding phase factors for $\gamma N\to\omega N$.
$\varphi$ stands for $\sigma$, $\pi$ and $f_2$. $f_2 NN$ coupling
constants are understood as $g^{(1)}_{f_2NN}=6.45$ and
$g^{(2)}_{f_2NN}=0$.}
    \begin{tabular}{c|c|c|c|c}\hline
        meson & trajectory($\alpha_\varphi$) & phase factor & $g_{\gamma \varphi \omega}$ & $g_{\varphi NN}$  \\
        \hline\hline
        $\pi$ for $\gamma p$ &  $0.7(t-m_{\pi}^2)$ & $e^{-i\pi \alpha_{\pi}}$ & $-0.69$ & $13.4$  \\%
        $\pi$ for $\gamma n$ &  $0.7(t-m_{\pi}^2)$ & $1$                      & $-0.69$ & $-13.4$  \\%
        $\sigma$ & $0.7(t-m_{\sigma}^2)$ & $(1+e^{-i\pi \alpha_{\sigma}})/2$ & $-0.17$ & 14.6 \\%
        $f_2$ &  $0.9(t-m_{f_2}^2)+2$ & $(1+e^{-i\pi \alpha_{f_2}})/2$ & 0.0376 & 6.45; 0.0 \\%
        $f_1$ &  $0.028\,t+0.9$ & $(-1+e^{-i\pi \alpha_{f_1}})/2$ & $0.18$ & 2.5 \\%
        \hline
        \hline
    \end{tabular}\label{tb1}
\end{table}

Together with the pseudoscalar $\pi$ exchange the exchange of
axial vector meson constitutes the contribution from the unnatural
parity. However, the case of $a_1$ exchange is excluded because it
does decay to neither $V\gamma$ nor $\gamma\gamma$ decay, while
the $f_1$ is possible to decay via both channels. As for the
reggeization of $f_1$, a special form of the trajectory
$\alpha_{f_1}(t)=0.028\,t+0.9$ is suggested in Ref.
\cite{kochelev}, which is deduced from QCD axial anomaly. In the
$\gamma p\to\phi \,p$ reaction we found it to contribute at high
energies with the flat slope and positive intercept comparable to
those of Pomeron \cite{kong2}. With an expectation of such a
special role at high energy, we include $f_1$ exchange in the
present model. We use the coupling constant $g_{f_1 NN}=2.5$
\cite{kong2}. However, since the decay width $f_1\to\omega\gamma$
is not measured yet, we refer to theoretical estimates such as the
constituent quark model \cite{ishida} where the partial decay
widths are estimated as $\Gamma_{f_1\to\rho^0\gamma}=509$ keV,
$\Gamma_{f_1\to\omega\gamma}=48$ keV, and
$\Gamma_{f_1\to\phi\gamma}=20$ keV, respectively. These are
comparable to the values 675 keV for the $f_1\to\rho^0\gamma$
estimated from the fraction
$(2.8\pm0.7)\times 10^{-2}$ \cite{amelin} to the full width
$\Gamma_{f_1}=24.1$ MeV, and 18.1 keV for the $f_1\to\phi\gamma$
reported in the PDG.
Thus, we adopt the estimate above to obtain $g_{\gamma
f_1\omega}=0.18$.  The nucleon axial form factor is taken the same
as in Ref. \cite{liesenfeld} with the cutoff mass $M_A=1.08$ GeV
for the sake of consistency.

Before proceeding, we give a few comments on the $t$-channel
exchanges. Here we neglect the pseudoscalar meson $\eta$ exchange
for simplicity, because its contribution is suppressed by the
larger mass and the smaller branching ratio $\omega\to\eta\gamma$
in comparison to $\pi$ exchange. For the spin-2 tensor meson,
$a_2$ meson could contribute as well in the isovector channel.
However, the determination of $g_{\gamma a_2\omega}$ is somewhat
confusing because the constituent quark model \cite{ishida}
predicts the decay width $\Gamma_{a_2\to\omega\gamma}=247$ keV
which leads to $g_{\gamma a_2\omega}=0.1$. It is larger than
$g_{\gamma f_2\omega}$ by a factor of 2.7, although the decay
$\Gamma_{a_2\to\gamma\gamma}=(9.4\pm0.7)\times10^{-6}$ is less
than $\Gamma_{f_2\to\gamma\gamma}=(1.64\pm0.19)\times10^{-5}$ by a
factor of 1.74 from the PDG. For a reasonable choice, in favor of
the two gamma decays based on the vector meson dominance, we
rather assume that the decay $\Gamma_{a_2\to\omega\gamma}$ is
similar to $\Gamma_{f_2\to\omega\gamma}$ in magnitude at best and
take the decay width $\Gamma_{a_2\to\omega\gamma}=26$ keV to
obtain $g_{\gamma a_2\omega}/m_0=0.033/m_0$. With
$g_{a_2NN}^{(1)}=1.4$ and $g_{a_2NN}^{(2)}=0$ chosen~\cite{bgyu},
its contribution turns out to be in minor role, which is less than
that of $f_2$ by a factor of $10^{-2}$ order. We, thus, exclude it
for simplicity.

We summarize the coupling constants and trajectories with the
corresponding phase factors in Table~\ref{tb1}.

\subsection{Regge cuts}

\begin{figure}[]
\centering \epsfig{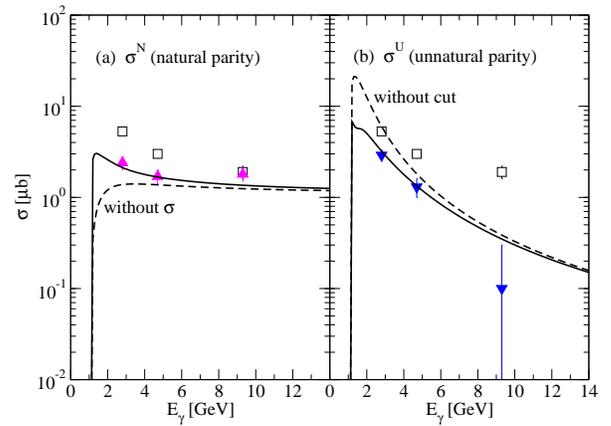}
\caption{Natural and unnatural parity cross sections, $\sigma^N$
and $\sigma^U$, for $\gamma p\to \omega p$.  In (a) solid and
dashed curves are with and without $\sigma$ meson exchange. In (b)
solid and dashed curves are with and without the cuts in the
unnatural parity cross section. Empty squares are the data of
total cross section. Data are taken from Ref.~\cite{ballam}. }
\label{fig1}
\end{figure}

Figure \ref{fig1} shows the  natural and unnatural parity cross
sections, $\sigma^N$ and $\sigma^U$, respectively \cite{ballam}.
In (a) the solid and dashed curves correspond to the cross section
with and without $\sigma$ exchange. This signifies the role of
$\sigma$ exchange indispensable as the peaking of cross section
$\sigma^N$ near threshold cannot be reproduced without it.

\begin{figure}[t]{}
\centering \epsfig{file=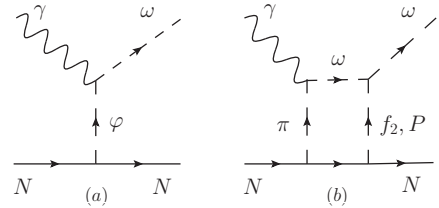, width=0.7\hsize}%
\caption[]{ Reggeons $\varphi=\pi,\,\sigma,\,f_2$ (a) and
absorptive cuts in the exclusive $\gamma N\to \omega N$ process at
forward angles (b). Elastic cuts $\pi$-$f_2$ and
$\pi$-$\mathbb{P}$ are viewed as the exchange of two Reggeons.}
\label{fig2}
\end{figure}

The case for the unnatural parity is shown in (b) where the dashed
curve is the cross section from the pure $\pi$ exchange. Thus,
without any sort of reduction, the exchange of $\pi$ Reggeon
should overestimate the data by a factor of two as shown in
Fig.~\ref{fig1}(b).  On the other hand, the contribution of $\eta$
meson exchange of the same parity is too small to improve the
overdominance of the $\pi$ exchange only by the coupling constants
$g_{\gamma\eta\omega}=0.161$ and $g_{\eta NN}=3.53$ \cite{titov}.
In order to modulate such a strong contribution of the $\pi$
exchange we need to introduce the Regge-cuts \cite{donnachie}
which are relevant to the Regge formulation of the production
amplitude rather than the form factors with cutoff masses. The
cuts are caused by absorption effects due to the elastic
scattering $\omega N\to\omega N$ through the sequential subprocess
as shown in Fig.~\ref{fig2}(b).

\begin{table}[t]
\caption{Cut parameters $C_{\varphi}$ and $d_\varphi$  given in
units GeV$^{-2}$. }
    \begin{tabular}{c|c|c|cc}\hline
        Reaction&Cut & $C_{\varphi}$ & $d_\varphi$&   \\
        \hline\hline
        $\gamma p$&$\pi$-$f_2$       &  $41$ & $2.2$ & \\%
                  &$\pi$-$\mathbb{P}$&  $-2.5$ & $2$ & \\%
\hline
        $\gamma n$&$\pi$-$f_2$       &  $11.5$ & $2.2$ & \\%
                  &$\pi$-$\mathbb{P}$&  $6$    & $2$ & \\%
        \hline
    \end{tabular}\label{tb2}
\end{table}

For the couplings allowed  for the  $\omega N\to\omega N$
subprocess in the elastic cut, isoscalar mesons $\sigma$ and $f_2$
in addition to the Pomeron exchange could be a candidate. Then,
the cuts are effective in the unnatural parity channel because the
overall naturality of the $t$-channel exchange in Fig. \ref{fig2}
(b) is unnatural by the product of pion naturality and the natural
parity exchanges of the mentioned mesons. In the calculation we
neglect $\sigma$ for the small intercept of $\pi$-$\sigma$ cut in
comparison to others, and write the $\pi$ Reggeon amplitude by
extending Eq. (\ref{pion}) to include the cuts as,
\begin{eqnarray}\label{cut1}
&&{\cal M}_{\pi}^{cut}=\widetilde{\cal M}_\pi\biggl[{\cal
R}^{\pi}(s,t)\nonumber\\
&&\hspace{1cm}+\sum_{\varphi=f_2,\mathbb{P}}C_{\varphi}\,e^{d_\varphi\,t}
e^{-i{\pi\over2}\alpha_{\pi\varphi}(t)} \left({s\over
s_0}\right)^{\alpha_{\pi\varphi}(t)-1}\,\biggr],\hspace{0.5cm}
\end{eqnarray}
with the $\varphi$ stands for $f_2$ and $\mathbb{P}$ in Fig.
\ref{fig2} (b). Here the $\widetilde{\cal M}_\pi$ denotes the pion
interaction vertices in Eq. (\ref{pion}) excluding the Reggeon
${\cal R}^\pi$. $C_\varphi$ is the strength of the cut  and
$d_\varphi$ is the parameter for the range of  the $\pi$-$\varphi$
cut to be fitted to data. Both parameters have the dimension of
GeV$^{-2}$.
The trajectory of the $\pi$-$\varphi$ cut in Eq. (\ref{cut1}) is
determined by the combination of $\pi$ and $\varphi$ meson
trajectories which is  given by
\begin{eqnarray}\label{cut2}
\alpha_{\pi\varphi}(t)=\alpha'_{\pi\varphi}\,t +
\alpha^0_{\pi\varphi}
\end{eqnarray}
with
\begin{eqnarray}\label{cut3}
\alpha'_{\pi\varphi}={\alpha^\prime_\pi\alpha^\prime_\varphi\over
\alpha^\prime_\pi+\alpha^\prime_\varphi}\,,\hspace{0.5cm}
\alpha^0_{\pi\varphi}=\alpha_\pi(0)+\alpha_\varphi(0)-1
\end{eqnarray}
for $\varphi=f_2$, $\mathbb{P}$, respectively. With the parameters
$C_\varphi$ and $d_\varphi$ in the cuts summarized in Table
\ref{tb2} we present the solid curve which agrees with the
unnatural parity cross section for $\gamma p$ reaction in Fig.
\ref{fig1} (b).  On the other hand, the $C_\varphi$ and
$d_\varphi$ for the $\gamma n$ Reaction in Table \ref{tb2}   are
fitted to the differential and total cross sections.

\section{results}

\subsection{$\gamma p\to\omega p$}

\subsubsection*{Differential and total cross sections}

\begin{figure}[]
\centering \epsfig{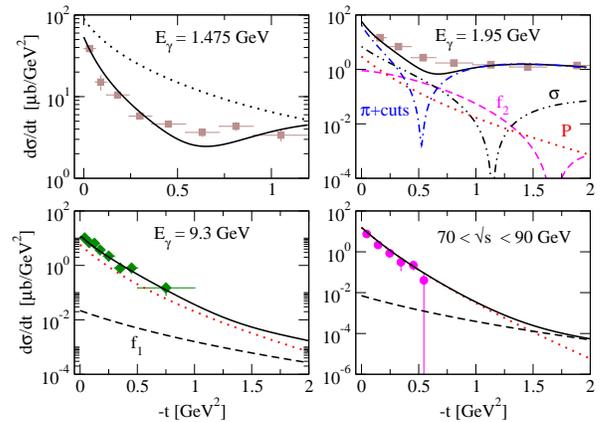}
\caption{Differential cross sections for $\gamma p\to \omega p$ in
four energy bins. Data are taken from Refs. \cite{dietz, ballam,
derrick}. Solid curves are from the full calculation. Dotted,
dashed, dash-dotted, and dash-dot-dotted curves are the
contributions of Pomeron, $f_2$, $\pi$+cuts, and $\sigma$ meson
exchanges, respectively.} \label{fig3}
\end{figure}

The $t$-dependence of differential cross sections are presented in
Fig. \ref{fig3} in four energy bins which are selected to
represent the energy region at threshold, intermediate, and high
energy, respectively. The overestimate of the single $\pi$
exchange without cuts is shown by the dotted curve at
$E_\gamma=1.475$ GeV. The contribution of $\pi$ exchange modulated
by the absorptive cuts gives a fair agreement with differential
cross sections. The role of $f_1$ exchange with the flat slope and
large intercept for the trajectory appears  to raise up the cross
section at high energies $70<\sqrt{s}<90$ GeV.

\begin{figure}[b]
\centering \epsfig{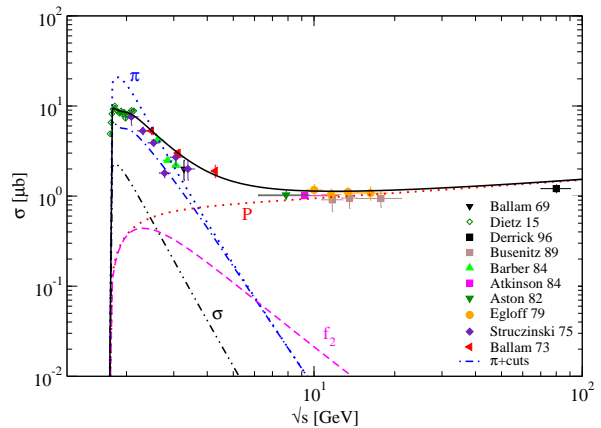}%
\caption{Total cross section for $\gamma p\to \omega p$. Data are
taken from Refs.
\cite{ballam,dietz,derrick,ballam69,busenitz,barber,atkinson,aston,
egloff,struczinski}. Notations for the curves are the same as in
Fig.~\ref{fig3}.} \label{fig4}
\end{figure}

Total cross section for $\gamma p\to\omega p$ is shown in Fig.
\ref{fig4} with the energy dependence from threshold
$\sqrt{s}\approx 1.72$ GeV  up to the realm of the Pomeron
exchange. At high energies, the Pomeron exchange is dominant and
the data are well reproduced by the single Pomeron exchange. Note
that the $\pi$ exchange without the cuts would make an
overestimate of the peak of the cross section near threshold. The
cross section in intermediate energies is sustained by the $f_2$
and Pomeron exchanges.

\subsubsection*{Density matrix and Beam polarization asymmetry}

\begin{figure}[]
\centering%
\epsfig{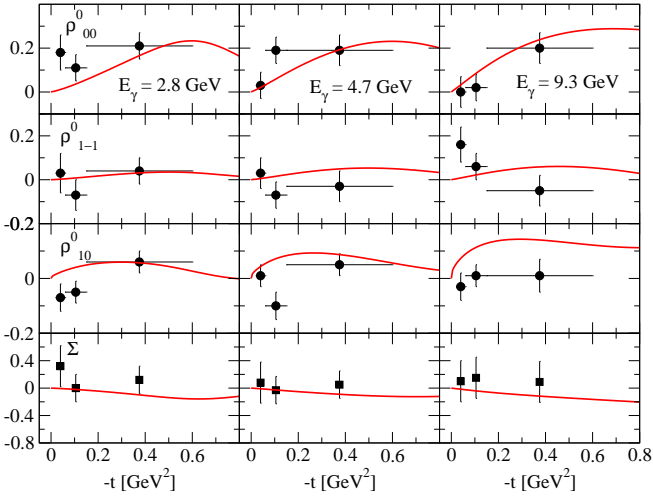}%
\caption{Spin-density matrix elements, parity and beam
polarization asymmetries in the Gottfried-Jackson frame for
$\gamma p\to \omega p$ at $E_\gamma=2.8$, 4.7 and 9.3 GeV. Data
are taken from Ref~\cite{ballam}. Beam polarization asymmetry is
obtained from spin density matrix elements via using Eq.
(\ref{eqzhao}).} \label{fig5}
\end{figure}

Spin density matrix elements $\rho^a_{ij}$ are important
observables because they are related with the single spin
polarizations such as the parity and beam polarization
asymmetries. The parity asymmetry $P_\sigma$ provides information
on the naturality of the reaction which is defined as
\begin{eqnarray}
P_\sigma=2\rho^1_{1-1}-\rho^{1}_{00}
\end{eqnarray}
in the Gottfried-Jackson frame. Furthermore, as the beam
polarization $\Sigma$ is one of important observables for the
study of recent JLab GlueX data we list it by extracting it from
the spin density matrix elements in Eq. (\ref{eqzhao}) below.

The beam polarization in the Gottfried-Jackson frame is defined as
\cite{zhao1}
\begin{eqnarray}
\Sigma=2\rho^1_{11}+\rho^{1}_{00}, \label{eqzhao}
\end{eqnarray}
with the trace condition $\sum_i \rho^0_{ii}=1$.
%
Figure \ref{fig5} shows a qualitative agreement of the spin
density matrix elements, parity asymmetry, and beam polarization
asymmetry with the data over the resonance region, $E_\gamma=2.8$,
4.7 and 9.3 GeV \cite{ballam}. We note that the prediction of
$P_\sigma$ is of the same quality as that of Ref. \cite{zhao1}.
The $P_\sigma\approx 1$ at $E_\gamma=9.3$ GeV implies that the
reaction proceeds via the natural parity exchange by its
definition $P_\sigma={d\sigma^N-d\sigma^U\over
d\sigma^N+d\sigma^U}$. The vanishing of $\Sigma$ over the
resonance region is a feature from the pure meson exchanges.
Nevertheless, our model fails to predict the beam polarization and
the double polarization at low energies such as beam-target in the
low energy region recently measured at the CBELSA/TAPS
\cite{dietz,eberhardt}. As pointed out in previous work
\cite{zhao} the contribution of  $N^*$ resonances in the
$s$-channel is essential in order to reproduce the observed spin
polarizations negative and nonvanishing at low energy, which is,
however, beyond the scope of the present work.

\subsubsection*{Scaling at large angle}

By the quark counting rule the energy dependence of differential
cross section obeys the power-law scaling, i.e.,
\begin{eqnarray}
s^{n-2}{d\sigma\over dt}\sim F(\theta)\,,
\end{eqnarray}
at the angle $\theta$  in the center of mass system. Thus, a
direct photon coupling  leads to $n=9$, whereas the vector meson
dominance needs $n=10$, respectively. These are expected to
exhibit a scaling either by the factor of $s^7$, or by $s^8$ as
the reaction energy increase. Therefore, a precise measurement of
the cross section around $\theta=90^\circ$ could provide us a
criterion to decide what portion of photon could convert to a
vector meson propagating as an effective degree of freedom.

For the saturation of a linear trajectory $\alpha(t)$ we use a
simple parameterization of the square root function \cite{collins}
\begin{eqnarray}\label{traj}
\alpha^*(t)=c_1+c_2\sqrt{t_1-t},
\end{eqnarray}
where the coefficients $c_1$ and $c_2$ are determined by the
boundary conditions $\alpha^*(t_0)=\alpha(t_0)$ and
$d\alpha^*(t_0)/dt=d\alpha(t_0)/dt$ at the saturation point $t_0$
where we choose to make the trajectory saturating to $-1$. Then,
$t_1$ is the initial point of the square root function the
argument of which should be positive and we take as $t_1>t_0$ in
the calculation.

\begin{figure}[]
\centering \epsfig{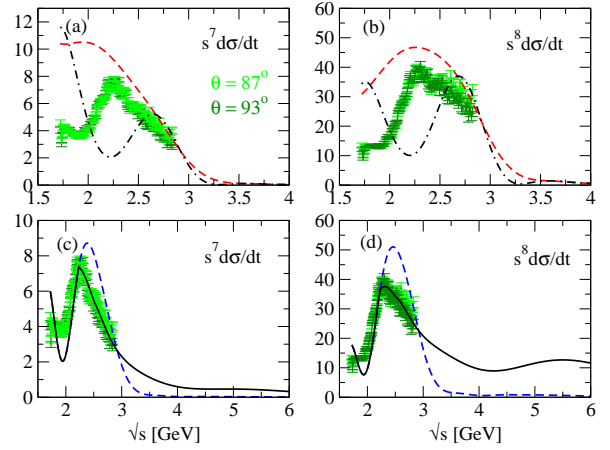}%
\caption{Differential cross sections for $\gamma p\to \omega p$ at
$\theta=90^\circ$ scaled by $s^7$ and $s^8$ in units of
$10^{7}$GeV$^{12}$nb and $10^{7}$GeV$^{14}$nb, respectively. By
using the linear trajectories the sensitivity of the cross section
to cuts as well as the phase of the $\pi$ exchange is investigated
in (a) for $s^7$ and (b) for $s^8$ scaled cross sections. Dashed
curves are from the full calculation where the complex phase is
taken for the $\pi$ exchange without cuts, whereas the dash-dotted
are from $\pi$ exchange with cuts but canonical phase,
$(1+e^{-i\pi\alpha_\pi})/2$, respectively. Nonlinearity of
trajectory is tested in (c) and (d). Given the $\pi$ exchange with
complex phase and cuts, solid curves result from trajectories of
$\pi$, $\sigma$, and $f_2$ saturated as shown in Fig. \ref{fig7}
(a), whereas the dashed ones are the results without saturation.
Data are taken from Ref.~\cite{williams}.} \label{fig6}
\end{figure}

\begin{figure}[b]
\centering \epsfig{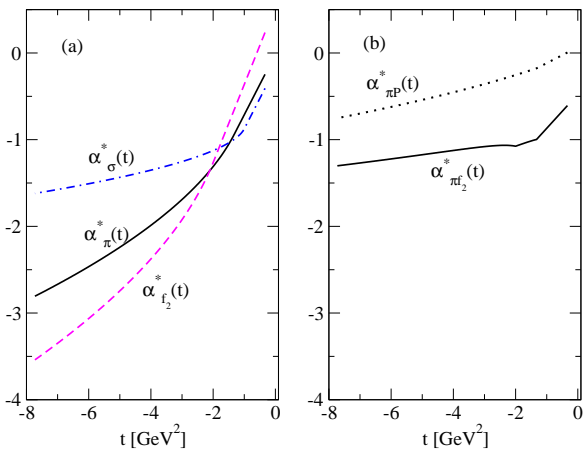}%
\caption{Saturation of trajectories for Reggeons (a) and cuts (b).
} \label{fig7}
\end{figure}

We now discuss the application of the present model to the scaled
cross sections at large $-t$. For doing this we consider the
saturation of a trajectory as $-t\to\infty$, in which case the
trajectory becomes independent of $t$, as a result
\cite{collins,sergeenko}.

\begin{figure}[]
\centering \epsfig{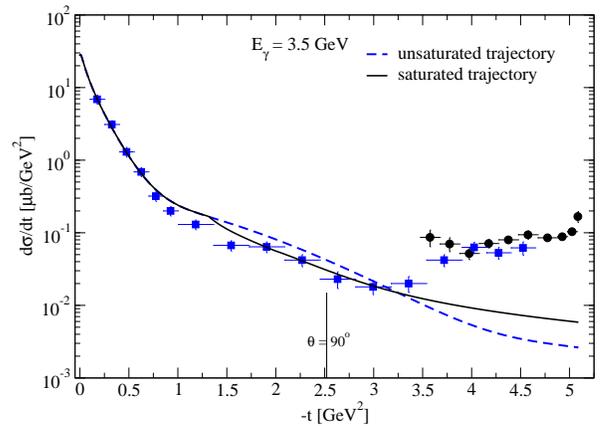}%
\caption{Differential cross section for $\gamma p\to\omega p$ as a
function of $-t$. Nonlinear trajectories agree with data around
$\theta=90^\circ$ with a limit of meson exchange model appearing
at the very backward angles.   Data are taken from Refs.
\cite{battaglieri,clifft}. } \label{fig8}
\end{figure}

For the sake of consistency, the nonlinear trajectory given by Eq.
(\ref{traj}) is applied to the cut trajectories for $\pi$-$f_2$
and $\pi$-$\mathbb{P}$ as well. Nevertheless, this does not expect
to affect the previous results because the cuts are by themselves
effective for the very small $-t$. The slopes and intercepts of
the $\pi$-$f_2$ cut trajectory in Eq. (\ref{cut2}) are, therefore,
parameterized as
\begin{eqnarray}\label{cut5}
\alpha'_{\pi f_2}={\alpha'^{*}_\pi\alpha'^{*}_{f_2}\over
\alpha'^{*}_\pi+\alpha'^{*}_{f_2}}\,,\hspace{0.2cm}%
\alpha^0_{\pi
f_2}=\alpha^{*}_\pi(0)+\alpha^{*}_{f_2}(0)-1,\hspace{0.3cm}
\end{eqnarray}
in Eq. (\ref{cut3}) with the starred quantities taken to be the
slope and intercept of Reggeon trajectory nonlinearized by Eq.
(\ref{traj}). However, in consideration of the characteristics of
the Pomeron exchange at the very forward angle we keep the Pomeron
trajectory unsaturated, while taking the nonlinear $\pi$
trajectory in the cut so that
\begin{eqnarray}\label{cut5}
\alpha'_{\pi\mathbb{P}}={\alpha'^{*}_\pi\alpha'_{\mathbb{P}}\over
\alpha'^{*}_\pi+\alpha'_{\mathbb{P}}}\,,\hspace{0.2cm}%
\alpha^{0}_{\pi
\mathbb{P}}=\alpha^{*}_\pi(0)+\alpha_{\mathbb{P}}(0)-1,\hspace{0.3cm}
\end{eqnarray}
for the $\pi$-$\mathbb{P}$ cut.

Figure \ref{fig6} shows predictions for scaled differential cross
sections for the $\gamma p$ reaction with the saturation of
trajectories by choosing $t_0=-1.325,\,t_1=-0.85$ for $\pi$, by
$t_0=-1.0,\,t_1=-0.95$ for $\sigma$, and by $t_0=-2.0,\,t_1=-1.5$
for  $f_2$ in unit of GeV$^2$. The data are collected from the
measurement at $\theta=87^\circ$ and $93^\circ$ by the CLAS
Collaboration \cite{williams} and re-sorted to the cross sections
$s^7d\sigma/dt$ and $s^8d\sigma/dt$, respectively. With the
trajectories unsaturated, we first examine the sensitivity of the
cross sections to the cuts and phase of $\pi$ exchange in (a) and
(b). In lower row (c) and (d) we note that the saturation of
trajectories at mid angle is significant and suggestive of
nonvanishing cross section over $\sqrt{s}\approx3$ GeV in the
$s^7$ and $s^8$ scaled data, respectively. In Fig. \ref{fig7}, we
show the nonlinear behavior of Reggeon and cut trajectories with
respect to momentum squared $t$. In order for an agreement with
data as shown in Fig. \ref{fig6} (c) and (d) we have to make the
saturation of the trajectories in Fig. \ref{fig7} (a) to have a
smooth decrease of the $t$ dependence rather than a strict
approach to an ideal limit $-1$.

The dependence of  differential cross section on the momentum
squared $-t$ is reproduced at $E_\gamma=3.5$ GeV with the
nonlinear trajectories and compared to the case without saturation
in Fig. \ref{fig8}. Around $\theta\approx 90^\circ$ the cross
section from the nonlinear trajectories is in good agreement with
existing data, though a limit of the $t$-channel Regge model to
the very backward angles $-t>3.5$ GeV$^2$ appears as a large
discrepancy with data. The rise of data over $-t\approx 3.5$
GeV$^2$ could be further accounted for by the $u$-channel nucleon
Reggeon, and a comprehensive description of the backward process
which could cover up this issue will appear elsewhere.

Before closing the application of the nonlinear trajectory to the
scaling of $\omega$ photoproduction at wide angles, a few remarks
should be in order. First, we observe that the absorptive cuts
play the role not only important to reduce the strength of the
$\pi$ exchange but also crucial to reproduce the scaled
differential cross section. At second, the canonical phase
$(1+e^{-i\pi\alpha_\pi})/2$ is not valid  to reproduce the scaled
cross section as shown in Fig.~\ref{fig6}, because it leads to a
minimum  at the nonsense zeros of the phase, i.e.,
$1+e^{-i\pi\alpha_\pi}=0$, which is opposite to the peak of the
cross section $s^7d\sigma/dt$ observed at $\sqrt{s}\simeq 2.27$.
Furthermore, as in Fig. \ref{fig6}(a) and (b), such a phase yields
the dips in the differential cross section which are not seen in
the experimental data. These support the complex phase for the
$\pi$ Reggeon as adopted in Table \ref{tb1}. Lastly, both the
$s^7$ and $s^8$ scaled cross sections in Fig. \ref{fig6}(c) and
(d) show nonzero cross sections $\approx 0.3$ and $10$ in their
respective sizes over $\sqrt{s}\approx3$ GeV, i.e., a $genuine$
scaling through the hard process. Otherwise, they are vanishing
with the linear trajectories. Moreover, noticing that the relative
size between $s^7$ and $s^8$ cross section is $\approx1/5$ below
$\sqrt{s}\approx 3$ GeV, and finding that it is further reduced to
$\approx 1/30$ in the scaling region as above, we expect that this
give us a clue to discern which one is probable between the direct
photon coupling and the vector meson dominance, if data exit
beyond $E_\gamma=3$ GeV further. Therefore, a measurement of the
cross section at $\theta=90^\circ$ extending to a region over
$\sqrt{s}\approx 3$ GeV is highly desirable.

\subsection{$\gamma n\to\omega n$}

\subsubsection*{Differential and total cross sections}

\begin{figure}[]
\centering \epsfig{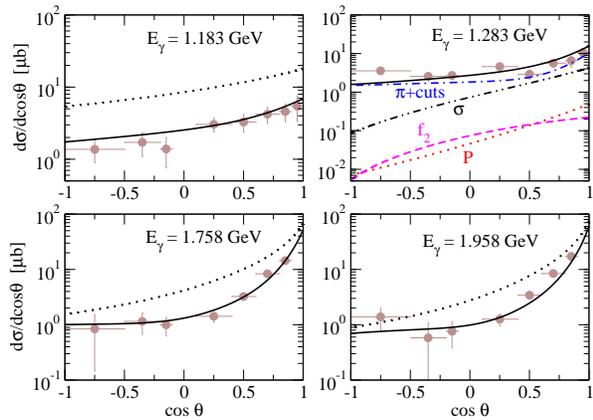}%
\caption{Differential cross sections for $\gamma n\to \omega n$.
Dotted curves are the cross sections without cuts. Dashed-dotted,
dash-dot-dotted, and dashed curves are from $\pi$ exchange with
cuts, $\sigma$, and $f_2$ exchanges, respectively. Pomeron
exchange is given by the red dotted curve. Data are taken from
Ref.~\cite{dietz}.} \label{fig9}
\end{figure}

\begin{figure}[ht]
\centering \epsfig{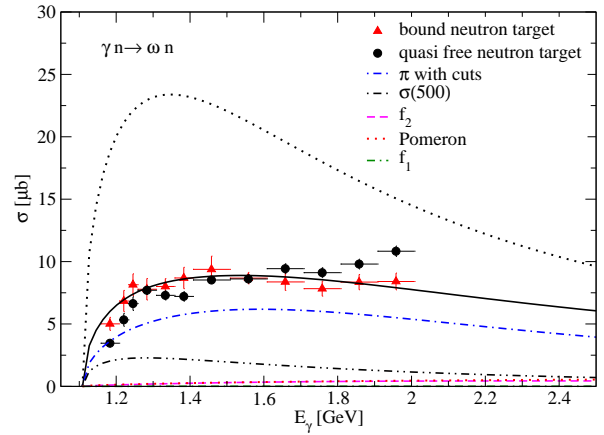}%
\bigskip
\caption{Total cross section for $\gamma n\to \omega n$. Solid
curve is the cross section from the full calculation with the cuts
included in the $\pi$ exchange. Cross section without the cuts is
shown by dotted curve which is wide out of data. Data are taken
from Ref.~\cite{dietz}.} \label{fig10}
\end{figure}

Experimental data on the $\gamma n$ reaction are not enough to
investigate the reaction mechanism up to high energy, and the data
recently measured at the CBELSA/TAPS Collaboration \cite{dietz}
are restricted only to a lower energy region. In this subsection
we calculate the energy and angle dependences of the $\gamma n$
reaction to provide  differential and total cross sections to
compare with data available.

In the photoproduction amplitude the only difference between
$\gamma p$ and $\gamma n$ reactions is the contribution of
isovector $\pi$ exchange with the sign of $g_{\pi nn}$ opposite to
$g_{\pi pp}$, as listed in Table \ref{tb1}. Within the present
framework we further modify the cut parameters and change the
phase of $\pi$ exchange to obtain an agreement with differential
and total cross sections.

Differential cross sections in the resonance region are reproduced
in Fig. \ref{fig9} where the constant phase of $\pi$ Reggeon is
taken to reproduce the CBELSA/TAPS data. The respective
contributions of meson exchanges are shown at $E_\gamma=1.283$
GeV.

Figure \ref{fig10} shows the total cross section for $\gamma n$
reaction at low energy with the data from the quasi-free neutron
and bound neutron in the deuteron targets, respectively. The cut
parameters in Table \ref{tb2} are fixed to reproduce the total
cross section of the reaction with the bound neutron in the final
state. As shown in the figure, the role of the cuts in the $\gamma
n$ reaction is crucial to agree with experiment as well.

\subsection{Application to $\omega\Delta$ photoproduction}

In this subsection we discuss the analysis of $\gamma
p\to\omega\Delta^+$ basically in the same framework as described
above for the $\omega$ photoproduction on nucleon, but the Dirac
spinor for the nucleon is replaced by the Rarita-Schwinger spinor
for the spin-3/2 $\Delta$ baryon in the final state. The
experimental data for the differential cross sections at
$E_\gamma=2.8$ - 4.8 GeV and the corresponding data points in the
total cross section were measured by the LAMP2 Group
\cite{barber}. More recent experiment of the reaction is performed
by the CB-ELSA Collaboration in which case the differential and
total cross sections were measured below $E_\gamma=3$ GeV
\cite{junkersfeld07}. In Ref. \cite{clark77} the Regge pole model
was applied for the $\gamma p\to\omega\Delta^+$  to discuss the
nondiffractive feature of the reaction with the conclusive results
from the role of the $\pi$ exchange primarily with cuts but the
tensor meson $a_2$ in minor role.

Here we pay our attention to the production mechanism rather in
the higher energy beyond $E_\gamma=3$ GeV because the data points
of the total cross section below the energy from the CB-ELSA
Collaboration are scattered. For the $\gamma p\to\omega\Delta^+$
process only the isovector exchange is allowed in the $t$-channel
by the transition of isospin 1/2 proton to the 3/2 $\Delta^+$.
Thus, neither the Pomeron nor the isoscalar mesons $\sigma$ and
$f_2$ are exchanged in the reaction process. In this work we
consider the $\pi$ exchange which is expected to give the leading
contribution with the large decay width $\omega\to\pi\gamma$. For
the test of the Regge pole fit to data \cite{clark77}, we also
include the $a_2$ exchange.

Pion exchange in the $t$-channel is gauge-invariant by itself with
the production amplitude given as,
\begin{eqnarray}\label{pion-delta}
&&{\cal M}_{\pi}=-i{g_{\gamma\pi\omega}\over m_0}{f_{\pi
N\Delta}\over m_\pi}\,\varepsilon_{\mu\nu\alpha\beta}\epsilon^\mu
\eta^{*\nu} k^\alpha Q^\beta\nonumber\\&&\hspace{1cm}\times
\overline{u}_\lambda(p')Q^\lambda u(p){\cal R}^\pi(s,t)
\end{eqnarray}
with the coupling constant $g_{\gamma\pi\omega}$ from Table
\ref{tb1}. The spin-3/2 spinor is denoted by $u_\lambda$.
The determination of the coupling constant $\pi$
exchange is of importance because it gives the most dominant
contribution to the process.   From the empirical value for the
maximum width $\Gamma_{\Delta\to\pi N}=120$ MeV taken from PDG,
the $\pi N\Delta$ coupling constant is estimated to be
$f_{\pi^-p\Delta^{++}} \approx 2.16$. According to the $\pi
N\Delta$ transition based on the quark model wave function, the
$\pi N\Delta$ coupling reads \cite{araki},
\begin{eqnarray}\label{pi3}
f_{\pi N \Delta}=\frac{6\sqrt{2}}{5}f_{\pi p p}
\end{eqnarray}
which gives $f_{\pi N\Delta} \approx 1.70$ with $f_{\pi p p}=1.0$
taken. Therefore, we consider $f_{\pi N\Delta}$ in the range
$1.7\leq f_{\pi N\Delta}\leq 2.16$ to take 1.7 for a consistency
with the quark model prediction.

For tensor meson $T(=a_2)$ exchange we assume a simple form for
the $TN\Delta$ from the $TNN$ vertex \cite{bgyu-pi-delta} in
consideration of parity and spin, i.e.,
\begin{eqnarray}
&&\overline{u}(p')(\gamma_{\sigma}P_\lambda+\gamma_{\lambda}P_\sigma)
u(p)e^{\sigma\lambda}\nonumber\\&&\hspace{1.5cm}
\to\overline{u}^{\nu}(p')(g_{\nu\sigma}P_\lambda+g_{\nu\lambda}P_\sigma)\gamma_5
u(p)e^{\sigma\lambda}
\end{eqnarray}
with $e^{\sigma\lambda}$ representing the spin polarization tensor
of tensor meson $a_2$.

The production amplitude reads
\begin{eqnarray}\label{amp4}
&&{\cal M}_{a_2}=\bar{u}^\lambda(p')\nonumber\\&&\times{4g_{\gamma
a_2\omega }\over m_{0}}\left[(k\cdot q
\epsilon_\beta-\epsilon\cdot q
k_\beta)\eta_\rho+(\epsilon\cdot\eta
k_\beta-k\cdot\eta\epsilon_\beta)q_\rho\right]\nonumber\\&&
\times\Pi^{\beta\rho;\sigma\lambda}(Q) {f_{a_2 N\Delta}\over
m_{a_2}}\overline{u}^{\nu}(p')\left(g_{\nu\sigma}P_\lambda
+g_{\nu\lambda}P_\sigma\right)\gamma_5 u(p){\cal
R}^{a_2},\nonumber\\
\end{eqnarray}
where $P=(p+p')/2$ and spin-2 projection
$\Pi^{\beta\rho;\sigma\lambda}(Q)$ given in Eq. (\ref{spin2}). We
use the trajectory $\alpha_{a_2}(t)=0.9\,(t-m_{a_2}^2)+2$ and the
coupling constant ${f_{a_2N\Delta}\over m_{a_2}}\approx -3
{f_{\rho N\Delta}\over m_\rho}$  following the $t$-channel
helicity Regge-pole fit to data in Ref. \cite{clark77,goldstein}.
For the relation above we adopt  $f_{\rho N\Delta}=8.57$
\cite{bgyu-rho-delta}. The radiative coupling constant is taken to
be $g_{\gamma a_2\omega}/m_0=0.033/m_0$, as discussed in Sec. II
A.

\begin{figure}[ht]
\centering \epsfig{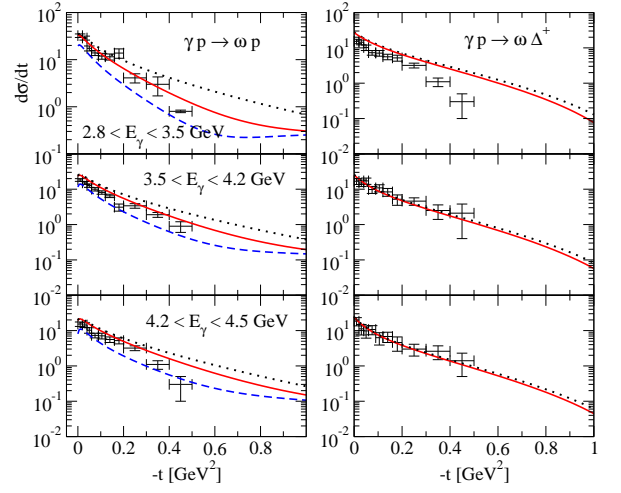}%
\bigskip
\caption{Comparison of differential cross sections between $\gamma
p\to\omega p$ and $\gamma p\to\omega \Delta^+$ photoproductions.
In both reactions the solid and dotted curves show the cross
sections with and without cuts, respectively. The dashed curve in
the left panel denotes the $\pi$ contribution with cuts. Data are
taken from Ref. \cite{barber}.} \label{fig11}
\end{figure}

\begin{figure}[h]
\centering \epsfig{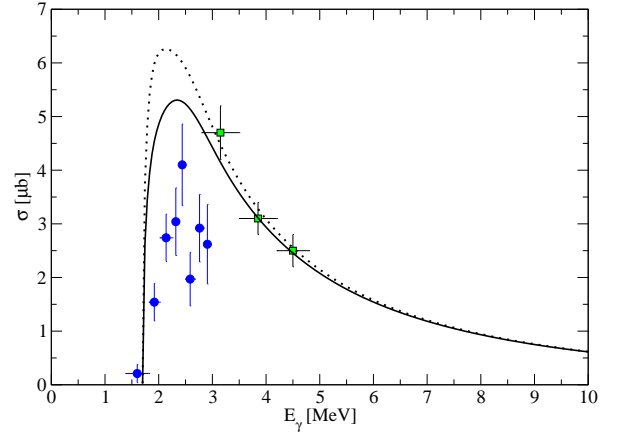}%
\bigskip
\caption{Total cross section for $\gamma p\to \omega \Delta^+$.
Solid and dotted curves are the cross sections with and  without
the cut. Data are taken from Refs. \cite{junkersfeld07,barber}.}
\label{fig12}
\end{figure}

In consideration of isospin relations \cite{clark77} we now write
the production amplitude as
\begin{eqnarray}\label{delta}
{\cal M}(\gamma p\to\omega\Delta^+)=\sqrt{2\over3}\left[ {\cal
M}^{cut}_\pi+{\cal M}_{a_2}\right]\,,
\end{eqnarray}
where the  ${\cal M}^{cut}_\pi$ includes the elastic cut of the
same form as in Eq. (\ref{cut1}) for the $\pi$-$a_2$ exchanges
with the cut parameters $C_{a_2}=4$ GeV$^{-2}$ and $d_{a_2}=1.5$
GeV$^{-2}$ chosen.

In the calculation the phase of the canonical form,
$(1+e^{-i\pi\alpha(t)})/2$, is taken for the $\pi$ and $a_2$
mesons in Eq. (\ref{delta}). Numerical consequences in the
differential cross sections are presented in Fig. \ref{fig11}
where the $\gamma p\to\omega\Delta^+$ reaction is compared with
the $\gamma p\to\omega p$ reaction. It is shown that most of the
contribution comes from the $\pi$ exchange in both processes.
However, in contrast to the $\gamma p\to \omega p$ reaction the
role of the $\pi$-$a_2$ cut is not quite clear in the case of
$\gamma p\to\omega\Delta^+$ as can be seen in the differential and
total cross sections in Figs. \ref{fig11} and \ref{fig12}. The
present model with the production amplitude in Eq. (\ref{delta})
predicts a good agreement with the differential cross section data
at $E_\gamma=8.9$ GeV in Ref. \cite{clark77}. Nevertheless in
order to clarify the cut effect on this reaction more data with
precision should be measured for the total cross section below
$E_\gamma=3$ GeV.  The maximum contributions of the $a_2(1320)$
exchange is found to be of $10^{-2}$ order in the total cross
section so that the differential and total cross sections are
completely reproduced by the $\pi$ exchange alone. The $a_2$
exchange found in minor role is consistent with the observation of
Ref. \cite{clark77}.

\section{summary and conclusions}

In summary we have investigated photoproduction of $\omega$ off a
nucleon target from threshold up to invariant energy
$\sqrt{s}\approx 100$ GeV within the Regge framework  for
$\sigma+f_2+\mathbb{P}+\pi+f_1$ exchanges.
The roles of these mesons are fixed by the consistency with the
natural and unnatural parity cross sections independently. Cross
sections for differential, total, spin density matrix and beam
polarization are reproduced to explain existing data on the
reaction $\gamma p\to \omega p$. Scaled differential cross
sections for $\gamma p\to \omega p$ are analyzed at the production
angle $\theta=90^\circ$ with the saturation of trajectories
substantial to agree with Jefferson Lab data. Differential and
total cross sections for the $\gamma n\to\omega n$ reaction are
analyzed to compare with recent experimental data  at the
CBELSA/TAPS Collaboration.
The most prominent feature of $\omega$ photoproduction at low
energies comes from the strong contribution of $\pi$ exchange
together with absorptive cuts. Need of  $\sigma$ meson exchange is
demonstrated in reproducing the natural parity cross section.
At high energies where the Pomeron exchange prevails, the $f_1$
exchange plays the role to raise the cross section in the large
$-t$ as expected from the trajectory specialized by the QCD
anomaly. But its role in the reaction at low energy is much
suppressed in comparison to others.
These characterize the features of $\gamma N\to\omega N$ reaction
for small $-t$, as viewed from the $t$-channel Reggeon exchange.
Analysis of scaled differential cross section is particularly
interesting, because it further supports the validity of the cuts
as well as the complex phase of the $\pi$ exchange both of which
are substantial to explain experimental data. Nonlinear
trajectories from a simple square root function are considered for
the saturation of the Reggeons and applied to describe the
reaction at large $-t$. Scaled cross sections by $s^7$ or $s^8$ at
$\theta=90^\circ$ are found to agree with existing data,
respectively, below $\sqrt{s}\approx 3$ GeV. Nevertheless, they
are predicted to behave different scaling over $\sqrt{s}\approx3$
GeV from each other, which could be a clue to distinguish the
priority between the direct photon coupling and vector meson
dominance. In this respect, we hope  that in future experiments
there should be a measurement of differential cross sections
around mid angle in the discussed region so that we could ask how
parton contributions arise there.

The leading role of the $\pi$ exchange continues at the $\gamma
p\to\omega\Delta^+$ reaction. However the role of the $\pi$-$a_2$
cut is found to be insignificant and the reaction is completely
dominated by the single $\pi$ exchange up to $E_\gamma=9$ GeV.
Therefore, it is concluded that the production mechanism of the
$\gamma p\to\omega\Delta^+$ process by the single $\pi$ exchange
of the unnatural parity is quite different from that of the
$\gamma p\to\omega p$ process in which case the isoscalar
$\sigma+f_2+$Pomeron exchanges of the natural parity play a role
up to high energies.

It is expected that the theoretical structures of the $\gamma
p\to\omega p$ and $\gamma p\to\omega\Delta^+$ photoproductions we
provided in this work will help understanding not only  the
reaction mechanisms themselves but also the vector meson
properties in the nuclear medium in future experiments.

       \section*{Acknowledgments}

This work was supported by the grant NRF-2017R1A2B4010117, and
partly by the grant NRF-2013M7A1A1075764 from National Research
Foundation (NRF) of Korea.

\end{document}